\documentclass{jpsj-suppl}
\usepackage{txfonts} 

\title{First-principles study of Exchange Interaction in Ising-type Multiferroic Ca$_{3}$CoMnO$_{6}$}

\author{Miho \textsc{Nishida}$^{1}$, Fumiyuki \textsc{Ishii}$^{2}$, and Mineo \textsc{Saito}$^{ 2}$}

\inst{
$^{\rm 1}$ Division of Mathematical and Physical Science, Natural Science and Technology, Kanazawa University, Kanazawa, 920-1192, Japan\\
$^{\rm 2}$ Faculty of Mathematics and Physics, Institute of Science and Engineering, Kanazawa University, Kanazawa, 920-1192, Japan\\
}

\email{m-nishida@cphys.s.kanazawa-u.ac.jp}

\recdate{September 30, 2013}

\abst{
We perform first-principles calculations of multiferroic Ca$_3$CoMnO$_6$ and 
evaluate the exchange coupling constants using the Green's function method. 
We clarify the effect of intra-chain and inter-chain exchange interactions 
on magnetic stability.
We find that inter-chain exchange coupling constants are antiferromagnetic and
that there are geometrical frustrations
in the triangular lattices of magnetic chains in Ca$_3$CoMnO$_6$.
The magnetic transition temperature is evaluated using
effective Hamiltonian with calculated exchange coupling constants.
We obtain the transition temperature 5.80K. 
The value has the same order as that of experimentally observed.
}

\kword{exchange coupling constants, Ising spin, triangular lattice, frustration, First-principles calculation, magnetic transition temperature}

\begin{document}
\maketitle

\section{Introduction}
Multiferroics having both ferromagnetic and ferroelectric properties attract wide scientific interests and are expected to be applied to spintronics devices. For examples, TbMnO$_3$, MnWO$_4$, Ni$_3$V$_2$O$_8$ and LiCuVO$_4$ have been studied\cite{KimuraTbMnO3,Taniguchi2006,Lawes2005,Park2007}. 
These materials have non-collinear antiferromagnetic spiral structures. 
Unlike these multiferroics, 
rare-earth free Ca$_3$CoMnO$_6$ has collinear Ising spin structure with strong anisotropy\cite{Y.J.Choi2008}.
Ca$_3$CoMnO$_6$ consists of triangular lattice and Ising chains expressed by axial next-nearest-neighbor Ising (ANNNI) model, 
and is thus, expected to have frustration of magnetic interaction.

Ca$_3$CoMnO$_6$ consists of spin chains in the hexagonal $c$ direction, which
form the triangular lattice in the $ab$ plane.
Neutron powder diffraction measurements clarified that the magnetic order of the chains  (Co-Mn-Co-Mn) in the ground state of Ca$_3$CoMnO$_6$ was $\uparrow \uparrow \downarrow \downarrow$\cite{Y.J.Choi2008}.
This order attracts scientific interests because of the  emergence of the electric polarization 
due to exchange striction mechanism. 

On the other hand, first-principles calculations predicted that the magnetic order was 
$\uparrow \downarrow \uparrow \downarrow$ \cite{Wu2009}. 
The inconsistency of stable magnetic order is expected to originate from the fact that, 
 previous theoretical calculations neglect inter-chain interactions.
In addition, if there are strong antiferromagnetic inter-chain interactions between triangular lattices, 
spin frustration is expected to be induced.

In this study, we clarify the stability of the spin structure of the chains. 
First, we perform first-principles calculations and evaluate exchange  
coupling constants using the Green's function method. 
Next, the effective Hamiltonian is obtained by using exchange coupling constants. 
Then, we evaluate the transition temperature of the magnetic phase transition.

\section{Computational Method}
By using the {\scriptsize OPENMX} code\cite{OpenMX}, we perform first-principles electronic-structure calculations 
based on the density functional theory (DFT) within the generalized gradient approximation (GGA)\cite{GGA-PBE}. 
The norm-conserving pseudopotential method\cite{PP-TM} is used. We use the linear combination of multiple pseudo 
atomic orbitals generated by a confinement scheme\cite{PAO1, PAO2}. 
Kohn-Sham orbitals are expressed 
$\psi_{\mu}({\bf r})  = \displaystyle{ \sum_{i \alpha} c_{\mu , i \alpha} \phi _{i \alpha} \left( {\bf r } - {\bf r }_{i} \right)} $ 
where, $i$ is a site index and $\phi _{i \alpha}$ is a numerical atomic orbital, where $\alpha = (plm)$ is an orbital index and $\phi _{i \alpha} \equiv Y_{lm}R_{ipl}$. A radial wave function $R_{ipl}$ depends on a site index $i$, a multiplicity index $p$ and an angular momentum quantum number $l$. 
The pseudo atomic orbitals are expanded Ca5.0-s2p2d2f1, Co5.5-s2p2d2, Mn5.5-s2p2d2, O4.0-s2p2d1.  
The former numbers (5.0, 5,5, 4.0) are the cutoff radii (a.u.) and the latter parts (s2p2d2f1 etc.) 
are the number of orbitals for  s, p, d and f composed.
The detail of formulation can be found in Ref.\cite{PAO2}.
The partial core correction\cite{PCC} is considered for all atoms. 
We use (4,4,4) uniform $\mbox{\boldmath $k$}$-point mesh for self-consistent calculations. 
The electron configurations of transition metal are considered to be high-spin states for Co$^{2+}$(d$^7$, $S$=3/2) 
and Mn$^{4+}$(d$^3$, $S$=3/2). The exchange coupling constants are calculated by using Green's function methods 
where magnetic force theorem\cite{Liechtenstein1987}. 
Applying magnetic force theorem to non-collinear magnetic perturbation for calculated ground state, we can obtain exchange interaction $J_{ij}$ between two different site $i$ and $j$ as following expression
$J_{ij} = \frac{1}{2\pi}\int^{\varepsilon _{F}} d \varepsilon \rm{Tr} [\hat{G}^{\uparrow}_{ij}\hat{V}_{j}\hat{G}^{\downarrow}_{ji}\hat{V}_{i} ]$
where, $\hat{G}^{\sigma}_{ij}$ is one particle spin-dependent Green's function constructed from Kohn-Sham orbitals and $\hat{V}_{i}$ is the on-site exchange interaction potential. The detail of formulation can be found in Ref.\cite{Han2004}.
The $\mbox{\boldmath $k$}$-point sampling for magnetic force 
theorem calculation is (4,4,4). We have confirmed that calculated exchange coupling constant varies very little 
if we doubled unit cell along the $c$-axis.

\section{Crystal Structure}
\begin{figure}
\begin{center}
\includegraphics[width=4cm]{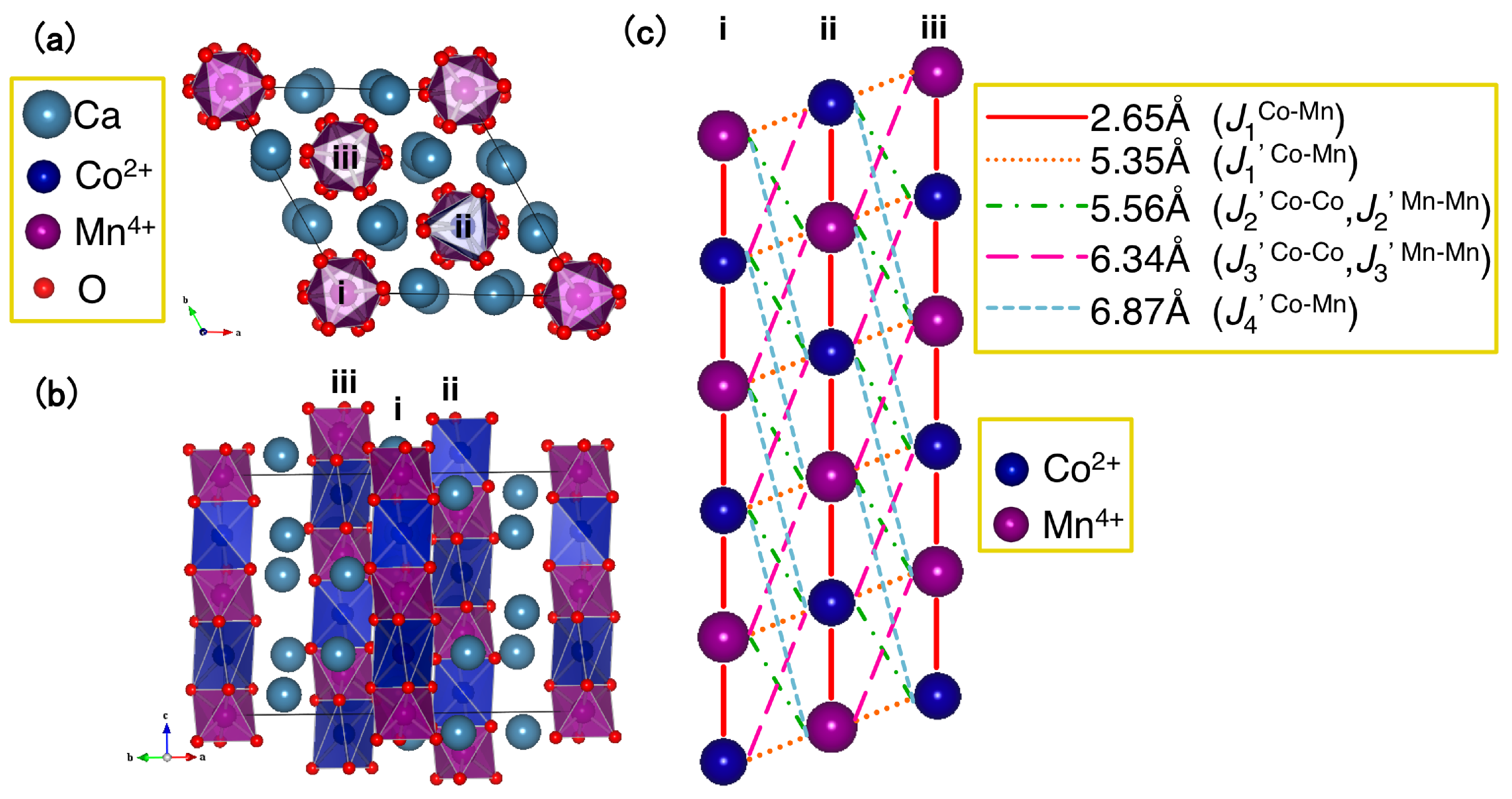}
\end{center}
\caption{(a) Crystal structure of [001] direction. (i), (ii) and (iii) represent quasi-one-dimensional Ising spin chains of (CoMnO$_6$)$_2$. (b) Crystal structure of [110] direction. (c) Atomic distances between magnetic atoms.}
\label{f1}
\end{figure}

The K$_4$CdCl$_6$-type Ca$_3$CoMnO$_6$ under the room temperature belongs to the space group $R\bar{3}c$ \cite{V.G.Zubkov2001}. The unit cell of the hexagonal type contains six formula units (66 atoms) and its perspective view is shown in Fig. \ref{f1}(a),(b). Experimental lattice constants are $a$=$b$= 9.1314 \AA, $c$ = 10.5817 \AA \cite{V.G.Zubkov2001}. Figure \ref{f1}(c) atomic distances in three chains. (i), (ii) and (iii) represent index of quasi-one-dimensional Ising spin chains of (CoMnO$_6$)$_2$. In Fig. \ref{f1}(a), we can see that the Ising spin chains are arranged in triangular lattices. Each chain consists of CoO$_6$ trigonal prisms and MnO$_6$ octahedra in $c$ direction. The former and latter are indicated by blue and magenta colors  in Fig. \ref{f1}(b). 
The oxygen atoms in the CoO$_6$ trigonal prisms and the MnO$_6$ octahedra are shared by Co and Mn.

\section{Results and Discussion}
\subsection{Evaluate of Exchange Coupling Constants}
\subsubsection{Intra-chain Exchange Coupling Constants}
In order to reveal the magnetic interaction in the intra-chain, 
we calculate exchange coupling constants between magnetic atoms using the Green's function method. 
We define the nearest neighbor exchange coupling constant as $J_1^{{\rm Co -Mn}}$, 
the next nearest neighbor exchange coupling constant between Co atoms as $J_2^{{\rm Co - Co}}$ 
and the next nearest neighbor exchange coupling constant between Mn atoms as $J_2^{{\rm Mn - Mn}}$ in intra-chain. 
We find that $J_1^{{\rm Co -Mn}}$ is larger than $J_2$, and $J_1^{{\rm Co -Mn}}$ is 
antiferromagnetic as shown in Table \ref{tb2}. 
Therefore, $\uparrow \downarrow \uparrow \downarrow$ order is stable. 
This result agrees with our first-principles total energy calculations and 
the previous first-principles calculation \cite{Wu2009}. 

\begin{table}[htbp]
\begin{center}
 \caption{Calculated intra-chain exchange coupling constants (meV).}
\begin{tabular}{c|c|c|c}
\hline
 Magnetic structure & $J_1^{{\rm Co -Mn}}$ & $J_2^{{\rm Co-Co}}$  & $J_2^{{\rm Mn-Mn}}$ \\
(distance \AA ) & (2.65) & (5.30) & (5.30) \\ \hline \hline
AFM($\uparrow \downarrow  \uparrow \downarrow$) & -2.15 & -0.08 & 0.19 \\ \hline
 \end{tabular}
 \label{tb2}
 \end{center}
 \end{table}

\subsubsection{Inter-chain Exchange Coupling Constants}
We calculate the inter-chain exchange coupling constants between magnetic atoms. 
As shown in \ref{f1}(c), 
we define the nearest neighbor exchange coupling constant as $J_1^{\prime {\rm Co-Mn}}$, 
the next nearest neighbor exchange coupling constant between Co atoms as $J_2^{\prime {\rm Co-Co}}$, 
the next nearest neighbor exchange coupling constant between Mn atoms as $J_2^{\prime {\rm Mn-Mn}}$, 
the third nearest neighbor exchange coupling constant between Co atoms as $J_3^{\prime {\rm Co-Co}}$, 
the third nearest neighbor exchange coupling constant between Mn atoms as $J_3^{\prime {\rm Mn-Mn}}$ 
and the fourth neighbor exchange coupling constant as $J_4^{\prime {\rm Co-Mn}}$ in inter-chain. 
As shown in Table \ref{tb3}, exchange coupling constants are antiferromagnetic except for $J_2^{'{\rm Co-Co}}$. 
In particular, the magnitude of $J_2^{\prime {\rm Mn-Mn}}$ is the largest exchange coupling constant in the inter-chain.
Furthermore, the magnitude of $J_2^{\prime {\rm Mn-Mn}}$ is larger than that of intra-chain next nearest neighbor 
$J_2^{{\rm Mn - Mn}}$. Therefore, it is necessary to consider inter-chain exchange coupling constants. 
\begin{table}[htbp]
 \begin{center}
  \caption{ Calculated inter-chain exchange coupling constants (meV). }
 \begin{tabular}{c|c|c|c|c|c|c}
 \hline
  Magnetic structure & $J_1^{'{\rm Co-Mn}}$ & $J_2^{'{\rm Co-Co}}$ & $J_2^{'{\rm Mn-Mn}}$ & $J_3^{'{\rm Co-Co}}$ &  $J_3^{'{\rm Mn-Mn}}$ & $J_4^{'{\rm Co-Mn}}$ \\
  (distance \AA ) & (5.35) & (5.56) & (5.56) & (6.34) & (6.34) & (6.87)  \\ \hline \hline
  AFM($\uparrow \downarrow  \uparrow \downarrow$) & -0.03 & 0.01 & -0.43 & -0.06 & -0.02 & -0.02 \\ \hline
 \end{tabular}
 \label{tb3}
 \end{center}
 \end{table}

\subsection{Effective Hamiltonian}
In order to study the energetics of all the possible spin 
configurations within the crystallographic unit cell, 
 we define the effective Hamiltonian as follows
\begin{eqnarray}
\mathcal{ H }_{eff} &=& \mathcal{ H }_{intra} + \mathcal{ H }_{inter},\label{eq:Ham} \\
\mathcal{ H }_{intra} &=& - \sum_{i<j}J^{{\rm Co-Mn}}_{1} \sigma _i \sigma _j -\sum_{k<l}J^{{\rm Co-Co}}_{2} \sigma _k \sigma _l -\sum_{k^{\prime}<l^{\prime}}J^{{\rm Mn-Mn}}_{2} \sigma _{k^{\prime}} \sigma _{l^{\prime}}, \\
\mathcal{ H }_{inter} &=& - \sum_{i<j}J^{\prime {\rm Co-Mn}}_{1} \sigma _i \sigma _j -\sum_{k<l}J^{\prime {\rm Co-Co}}_{2} \sigma _k \sigma _l -\sum_{k^{\prime}<l^{\prime}}J^{\prime {\rm Mn-Mn}} _{2}\sigma _{k^{\prime}} \sigma _{l^{\prime}} \nonumber \\
&-& \sum_{m<n}J^{\prime {\rm Co-Co}}_{3} \sigma _m \sigma _n -\sum_{m^{\prime}<n^{\prime}}J^{\prime {\rm Mn-Mn}} _{3}\sigma _{m^{\prime}} \sigma _{n^{\prime}} - \sum_{i^{\prime}<j^{\prime}}J^{\prime {\rm Co-Mn}}_{4} \sigma _{i^{\prime}} \sigma _{j^{\prime}},\label{eq:Ham1}
\end{eqnarray}
where $\sigma$ denote $\pm 1$, i.e., classical Ising spin.

\subsection{Total Energy of the Magnetic Structure}
\begin{figure}
\begin{center}
\includegraphics[width=3cm]{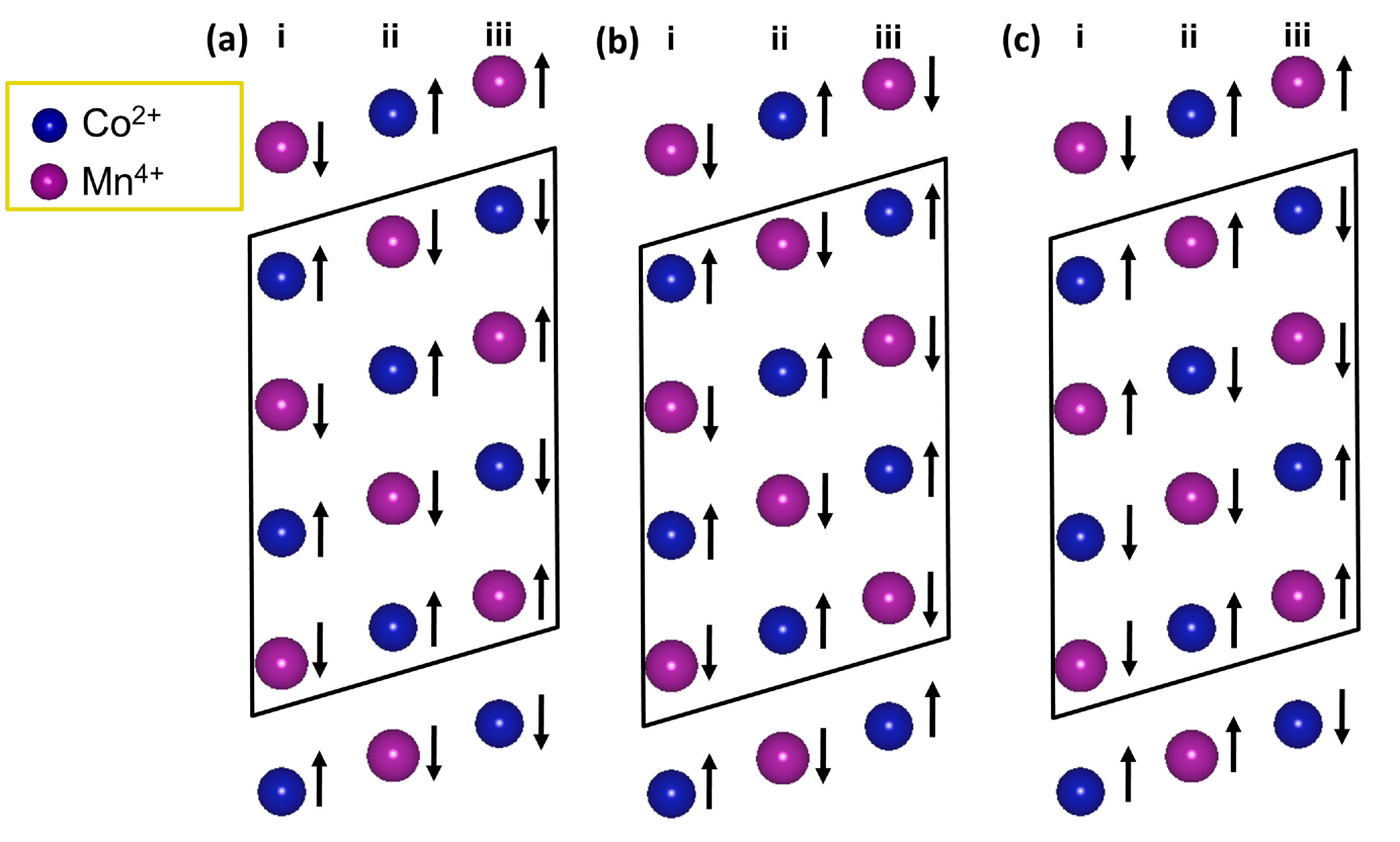}
\end{center}
\caption{(a) The magnetic structure of the ground state calculated by the effective Hamiltonian. 
(b) The chain (iii) is shifted to next site compared to Fig.\ref{f8}(a).
(c) The experimentally observed structure. The black box is the unit cell of Ca$_3$CoMnO$_6$.}
\label{f8}
\end{figure}

To evaluate the ground state, we consider triangular lattices of the spin chains in the crystallographic unit cell. Each spin chain consists of the periodic four magnetic atoms (Co-Mn-Co-Mn). Then, there are 12 magnetic atoms in the unit cell.
By using the above Hamiltonian eqs.(\ref{eq:Ham}) - (\ref{eq:Ham1}), we calculate the total energy (T=0) of 4096($=2^{12}$) spin configurations which are all possible spin configurations in the crystallographic unit cell. 
In the most stable structure, the spin configurations three chains are $\uparrow \downarrow \uparrow \downarrow$ as shown in Fig.\ref{f8}(a). 
This result agrees with our first-principles total energy calculations.
Fig.\ref{f8}(b) is the magnetic structure where the chain (iii) is shifted to next site compared to Fig.\ref{f8}(a).
Total energy of Fig.\ref{f8}(b) becomes +3.16 meV/f.u. higher energy than Fig.\ref{f8}(a). 
This is due to the strong interchain exchange interactions.

\begin{figure}
\begin{center}
\includegraphics[width=8.2cm]{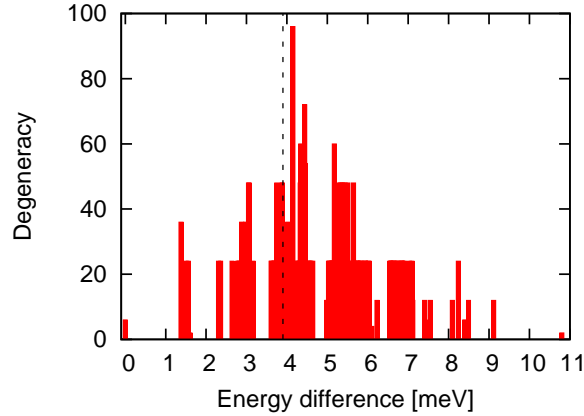}
\end{center}
\caption{The energy difference vs. the degeneracy factor. The vertical axis is the number of the magnetic structure comprising the same energy difference. The dotted line shows the energy of experimentally observed spin configuration (Fig.\ref{f8}(c)).} 
\label{f7}
\end{figure}

As for the $\uparrow \uparrow \downarrow \downarrow$ spin configurations,  
Fig.\ref{f8}(c) is experimentally observed magnetic structure. 
The magnetic order $\uparrow \uparrow \downarrow \downarrow$(Fig.\ref{f8}(c)) 
has +3.90 meV/f.u. higher energy than the most stable magnetic structure (Fig.\ref{f8}(a)). 
Due to the strong interchain exchange interactions, this $\uparrow \uparrow \downarrow \downarrow$ 
spin configurations in Fig.\ref{f8}(c) also change there energies  
decrease/increase up to -0.28meV/f.u./+1.64meV/f.u. 
if we shift the spin along the chain.
Figure \ref{f7} shows, density of magnetic states, degeneracy of magnetic states on total energy differences 
obtained from eqs.(\ref{eq:Ham}) - (\ref{eq:Ham1}) in all magnetic structures. 
4096 spin configurations are classified into 187 groups by energy and symmetry. 
We find that the energy distribution of 4096 spin configurations in narrow. 
In particular, the energies of 1094 spin configurations are less than the energy of experimentally observed spin configuration, 3.90meV/f.u., which is indicated by a dotted line in Fig.\ref{f7}. 
This is because the magnetic frustration in the triangular lattice of spin chains due to antiferromagnetic coupling. 
This feature may be the origin of 
{\it untrue long-range order nature} reported by
neutron diffraction measurements in Ca$_3$CoMnO$_6$\cite{Y.J.Choi2008}.

\subsection{Transition Temperature with Exact Enumeration}
\begin{figure}
\begin{center}
\includegraphics[width=8cm]{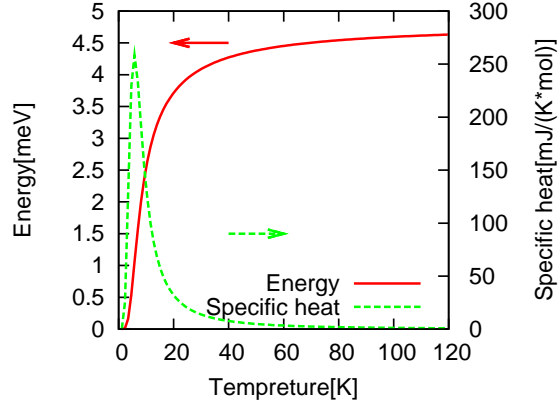}
\end{center}
\caption{The energy expectation and the specific heat by the partition function with exact enumeration. The maximum point of the specific heat is 5.80K.}
\label{f6}
\end{figure}

The method of exact enumeration in statistical physics is a typical method to evaluate the partition function of the simple model \cite{H.Gould}. 
To evaluate the transition temperature, we first calculate the partition function $Z$ by using calculated total energy differences. Then, we calculate expectation value of  energy $\langle E \rangle$ and specific heat $\langle C \rangle$.
The formulas are $Z = \displaystyle {\sum^{4096}_{i} \exp\left( - \frac{E_{i}}{k_{B}T} \right)}$, $\langle E \rangle = \displaystyle {\sum^{4096}_{i} \frac{E_{i} \exp\left( - \frac{E_{i}}{k_{B}T} \right)}{Z}}$ and 
$\langle C \rangle = \frac{\langle E^{2} \rangle - \langle E \rangle ^{2}}{k_{B}T}$, where $k_{B}$ is Boltzmann constant and $T$ is temperature.
The results are shown in Fig.\ref{f6}. Then we estimate magnetic transition temperature from result of calculated specific heat. 
We estimate the transition temperature by 
finding the maximum point of the specific heat.
The calculated magnetic transition temperature T$_c$ = 5.80 K is the same order of magnitude 
with experimental T$_c$ = 16.5K\cite{Y.J.Choi2008}. 

\section{Summary}
We perform first-principles calculations of Ca$_3$CoMnO$_6$
and evaluate the exchange coupling constants using the Green's function method. We clarify that 
considering inter-chain exchange coupling constants are substantial. 
The inter-chain exchange coupling constants are mostly antiferromagnetic and the largest one exceeds second nearest neighbor 
intra-chain exchange interaction. 
This interchain antiferromagnetic coupling 
causes the frustration in the triangular lattice of spin chain.
We calculate 4096 total energies using 
Ising Hamiltonian with the exchange coupling constants. 
We find that a large number of magnetic structures distribute in a narrow energy range. 
This is due to the frustration in triangular lattice of spin chain 
induced by antiferromagnetic interchain coupling.
We estimate the transition temperature by using the partition function and obtain T$_c$ = 5.80 K. 
The  magnetic transition temperature has the same order as that of experimentally observed T$_c$ (16.5 K).
Nevertheless, the ground state in our result is not agreement with experimentally result. Both the ground state and the transition temperature depend on the exchange coupling constants. Hence, we try other methods to include the exchange correlation effect 
such as LDA+U, LDA+DMFT \cite{Han2004, Liechtenstein2001} which are more reliable than the GGA to obtain exchange coupling constants.
We also should confirm the convergence the cell-size dependence of the exchange coupling constants when we use magnetic force theorem. 

\section{Acknowledgment}
Part of this research has been funded by the MEXT HPCI Strategic Program.
This work was partly supported by Grants-in-Aid for Scientific Research
(Nos. 25104714, 25790007, and 25390008) from the JSPS.
The computations in this research were performed using
the supercomputers at the ISSP, the University of Tokyo.


\end{document}